\documentclass[fleqn,usenatbib]{mnras}

\usepackage{newtxtext}
\usepackage{mathptmx}
\usepackage{comment}

\usepackage[T1]{fontenc}

\DeclareRobustCommand{\VAN}[3]{#2}
\let\VANthebibliography\thebibliography
\def\thebibliography{\DeclareRobustCommand{\VAN}[3]{##3}\VANthebibliography}

\usepackage{graphicx}	
\usepackage{amssymb}	
\usepackage{amsmath}	

\def\apj{ApJ}           	
\def\apjl{ApJ}          	
\def\apjs{ApJS}         	
\def\aap{A\&A}          	
\def\mnras{MNRAS}       	

\title[Fragmentation of ring galaxies]{Fragmentation of ring galaxies and transformation to clumpy galaxies}

\author[S. Inoue, N. Yoshida \& L. Hernquist]{
{Shigeki Inoue$^{1,2}$\thanks{E-mail: inouesg@ccs.tsukuba.ac.jp}, Naoki Yoshida$^{3,4,5}$ \& Lars Hernquist$^6$}
\\
$^{1}$Center for Computational Sciences, University of Tsukuba, Ten-nodai, 1-1-1 Tsukuba, Ibaraki 305-8577, Japan\\
$^{2}$Chile Observatory, National Astronomical Observatory of Japan, Mitaka, Tokyo 181-8588, Japan\\
$^{3}$Kavli Institute for the Physics and Mathematics of the Universe (WPI), UTIAS, The University of Tokyo, Chiba 277-8583, Japan\\
$^{4}$Department of Physics, School of Science, The University of Tokyo, Bunkyo, Tokyo 113-0033, Japan\\
$^{5}$Research Center for the Early Universe, School of Science, The University of Tokyo, Bunkyo, Tokyo 113-0033, Japan,\\
$^{6}$Institute for Theory and Computation, Harvard-Smithsonian Center for Astrophysics, 60 Garden Street, Cambridge, MA 02138, USA
}

\date{Accepted XXX. Received YYY; in original form ZZZ}

\pubyear{2020}

\begin{document}
\label{firstpage}
\pagerange{\pageref{firstpage}--\pageref{lastpage}}
\maketitle

\begin{abstract}
We study the fragmentation of collisional ring galaxies (CRGs) using a linear perturbation analysis that computes the physical conditions of gravitational instability, as determined by the balance of self-gravity of the ring against pressure and Coriolis forces.  We adopt our formalism to simulations of CRGs and show that the analysis can accurately characterise the stability and onset of fragmentation, although the linear theory appears to under-predict the number of fragments of an unstable CRG by a factor of 2. In addition, since the orthodox `density-wave' model is inapplicable to such self-gravitating rings, we devise a simple approach that describes the rings propagating as material waves. We find that the toy model can predict whether the simulated CRGs fragment or not using information from their pre-collision states. We also apply our instability analysis to a CRG discovered at a high redshift, $z=2.19$. We find that a quite high velocity dispersion is required for the stability of the ring, and therefore the CRG should be unstable to ring fragmentation. CRGs are rarely observed at high redshifts, and this may be because CRGs are usually too faint. Since the fragmentation can induce active star formation and make the ring bright enough to observe, the instability could explain this rarity. An unstable CRG fragments into massive clumps retaining the initial disc rotation, and thus it would evolve into a clumpy galaxy with a low surface density in an inter-clump region.

\end{abstract}




\section{Introduction}
\label{Intro}
A collisional ring galaxy (CRG) is formed by a head-on merger between a disc galaxy and a companion that penetrates a region near the disc centre \citep[e.g.][]{ts:76}. The formation of ring structures is driven by radial perturbations in the disc plane, and they propagate outwards. Their formation, evolution and details of the ring structures have long been investigated in a number of theoretical and observational studies, including $N$-body, hydrodynamical and cosmological simulations \citep[e.g.][]{ts:77,hw:93,mmr:08,raa:18}.   

\citet{lt:76} have proposed an analytical model that describes the formation and propagation of rings as `density waves' resulting from crowding of stars in a disc. If the merger of a companion is impulsive and vertically passes through the disc centre, the disc stars acquire velocities inwards. The head-on collision thus induces epicyclic motions of the stars \citep{bt:08}. Their epicyclic frequencies $\kappa$ generally decrease with radius. Therefore, when the stars at a given radius rebound and move outwards, other stars at outer radii are still moving inwards. Their collective motions generate a high-density ring region between these radii, and the stellar ring propagates outwards. Note that the stars do not migrate along with the ring propagation. This theory is thought to be valid when the self-gravity of the ring is negligible and the gas mass in the disc is small \citep[e.g.][]{s:10}.

Ring structures can, however, be self-gravitating if they are massive. The simulations of \citet{hw:93} demonstrate that CRGs can fragment into several clumps when head-on mergers drive strong perturbations that can sweep up a large amount of gas and stars into the ring. Although analytical studies argue that dynamical instability can lead a ring structure to collapse locally \citep[e.g.][]{r:42,w:74}, none of them has examined whether such an instability analysis can explain the fragmentation of CRGs. 

\begin{figure*}
  \includegraphics[bb=0 0 1540 384, width=\hsize]{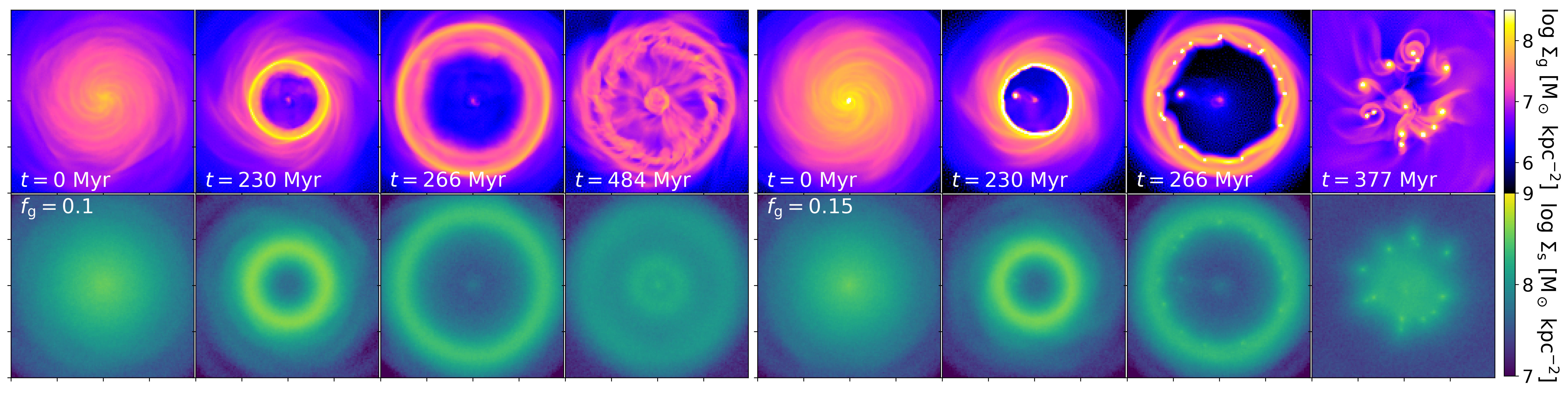}
\caption{Face-on density distributions of the gas (top) and stellar (bottom) components in our CRG simulations. The left and right sets of panels indicate the runs with $f_{\rm g}=0.1$ and $0.15$. The side length of each panel corresponds to $20~{\rm kpc}$. In both runs, the companions pass through the disc centres at $t=193~{\rm Myr}$.}
\label{snaps}
\end{figure*}
In this study, we perform simulations of CRGs in Section \ref{sims} and apply our linear perturbation analysis presented in Section \ref{ana} to the simulation results. Our analysis has been successfully applied to the (in)stability of spiral arms \citep{iy:18,iy:18b,iy:20,itm:21}, and it is also well-suited for the dynamical analysis of ring structures. We demonstrate that our formalism can characterise the dynamical (in)stability of the rings in our simulations. 

The propagation of such self-gravitating rings formed by strong perturbations is beyond the applicable domain of the density-wave theory of \citet{lt:76}. In Section \ref{Sweep}, we consider a simple model to describe the formation of a massive ring in the limit of strong perturbations, in which the impact of a merger is assumed to accumulate all gas and stars of the disc into the ring.

In Section \ref{highz}, we discuss a high-redshift CRG, R5519, found at $z=2.19$ \citep{yel:20}. We apply our analysis to this CRG and deduce its dynamical state. From our result, we argue about the origin of the high surface brightness of R5519 and speculate possible reasons why CRGs are rarely found at high redshifts. We present discussion and our conclusions in Section \ref{discussion}.

\section{Simulations}
\label{sims}
We use the moving-mesh code \textsc{Arepo} \citep{arepo,wsp:20} to follow the formation and evolution of CRGs, in which a galactic disc is modelled as $10^6$ gas cells and $5\times10^6$ stellar particles in a rigid halo potential. Our simulation models are similar to the fiducial example of \citet{hw:93} that produces a CRG closely resembling the Cartwheel galaxy. Our initial conditions (ICs) assume a galaxy having an exponential density profile with a total disc mass of $5.6\times10^{10}~{\rm M_\odot}$ and a scale radius (height) of $3.5$ ($0.7$) ${\rm kpc}$. The galaxy has no bulge and is embedded within a halo represented by a fixed isothermal potential with a total mass of $1.6\times10^{11}~{\rm M_\odot}$, a core radius of $3.5~{\rm kpc}$ and a cut-off radius of $35~{\rm kpc}$. Unlike \citet{hw:93}, the halo and a companion galaxy are represented with rigid potentials. The equation of state of the gas is isothermal, and the temperature is maintained at $T=10^4~{\rm K}$. In \textit{this} IC, the gas fraction of the disc is set to $f_{\rm g}=0.1$. The softening length of the stellar particles is chosen to be $\epsilon_{\rm s}=50~{\rm pc}$, and that of the gas cells is $\max(\epsilon_{\rm s}, 2.5r_{\rm g})$, where $r_{\rm g}$ is an approximate cell size. The initial velocity distribution is determined using the method of \citet{h:93} with the minimum $Q$ value of $1.3$, and the stellar and gas components in the disc initially have the same spatial and velocity distribution.

The above ICs are, however, not in dynamical equilibrium, and transient ring structures emerge within a few rotation periods even in the absence of a merger. We, therefore, evolve the disc in isolation for $1~{\rm Gyr}$. During this time, the transient rings expand and disappear while propagating through the outer regions. After this procedure, the disc does not form such a ring. We use the last snapshot of this procedure as the actual ICs of our CRG simulations; we refer to these as `relaxed ICs'. We generate two sets of the relaxed ICs. In the first one, the gas fraction is kept constant at $f_{\rm g}=0.1$ during the relaxing procedure. In the second one, the gas fraction gradually increases at a constant rate for $1~{\rm Gyr}$ and reaches $f_{\rm g}=0.15$ while the total disc mass is unchanged. In both cases, the gas and stellar density distributions in the relaxed ICs are hardly different from the initial exponential profile. The velocity dispersion of the stars is also unchanged, but that of the gas decreases to values significantly lower than the sound speed. Therefore, the gas discs become thinner than $0.7~{\rm kpc}$ in the relaxed ICs.

We let a companion galaxy collide with the disc of the relaxed ICs. The companion is represented by a spherical potential of \citet{h:90}, where the total mass is $5.6\times10^{10}~{\rm M_\odot}$, and the scale radius is $1.75~{\rm kpc}$. It is initially placed at $52.5~{\rm kpc}$ above the disc centre and on a straight orbit towards the centre with the escape velocity of $186~{\rm km~s^{-1}}$. The trajectory of the companion is determined by the gravity of the disc and the halo in a self-consistent manner. We run two simulations with the relaxed ICs with $f_{\rm g}=0.1$ and $0.15$ and the same companions. We also perform another set of similar simulations with different ICs in Appendix \ref{App}.

Various simplifications and arbitrary parameter settings have been made in our simulations. We aim to extract the key physics of the fragmentation of a CRG by comparing our analytical prediction (Sections \ref{ana} and \ref{Sweep}) with the simulations. We therefore perform our simulations under the conditions closest to the analysis, and our simulations include neither gas cooling, star formation nor stellar feedback. The isothermal assumption for gas is for consistency with the analysis where gas is assumed to be barotropic.\footnote{\citet{iy:20} have shown that the instability analysis can capture the fragmentation of spiral arms even in cosmological simulations including gas cooling.} The temperature $T=10^4~{\rm K}$ corresponds to the value to which atomic cooling by hydrogen and helium can cool gas down, and it is higher than the typical temperature of dense gas that forms molecules in a disc galaxy. However, cosmological simulations including metal-line cooling and feedback show that a significant fraction of disc gas in inter-arm regions is nearly isothermal at $T\sim10^4~{\rm K}$ \citep{iy:19}. The disc scale height of $h=0.7~{\rm kpc}$ may seem to be somewhat large if comparing with local spiral galaxies such as the Galactic thin disc with $h\sim0.3~{\rm kpc}$ \citep[e.g.][]{jib:08}. It is, however, not unrealistic for the arguments in Section \ref{highz} since high-redshift discs are generally thicker \citep[e.g.][]{ee:06}. In the simulations, the halo and the merging companion are represented as rigid potentials. Because the merging companion can be disrupted to some extent, it needs to be more massive than in our simulations to give the same impact to a disc. The halo of the disc galaxy is also affected by the merging. The density distribution of the halo would actually differ from that assumed in the simulations when a ring forms. However, since our analysis does not need to know a halo potential and is flexible to its change, assuming the rigid potential for a halo does not affect our conclusions.

Fig. \ref{snaps} shows the gas and stellar distributions in the two simulations. In the case of $f_{\rm g}=0.1$ (the left panels), the first ring is quickly formed but does not fragment. It expands and dissipates at $R\simeq10~{\rm kpc}$. A certain amount of gas falls back to regions around the centre, and a few successive rings form recursively. Behind the second ring, a number of `spokes' are found in the gas distribution (see the snapshot at $t=484~{\rm Myr}$ in Fig. \ref{snaps}). In the case of $f_{\rm g}=0.15$ (the right panels), knotty structures form in the first ring when it is located at $R\simeq5~{\rm kpc}$. The ring fragments into several massive clumps. The remnant retains the disc rotation and therefore would be observed as a clumpy galaxy, and the clumps keep orbiting around $R\sim3$--$8~{\rm kpc}$ (i.e. a few times the disc scale length). These orbital radii are consistent with galactocentric distances of observed clumps \citep[e.g.][]{tkt:13II,sok:16}. At $t=377~{\rm Myr}$, the number of the clumps is $N_{\rm clump}=13$ in the galaxy, whose mean gas mass is $3\times10^8~{\rm M_\odot}$. Although quantitative comparison requires more realistic simulations, the masses of the simulated clumps are consistent with the typical masses of observed clumps \citep[e.g.][]{grb:18}. The number of clumps, $N_{\rm clump}=13$, in the fragmenting CRG may be more than those of typical clumpy galaxies at high redshifts \citep[$N_{\rm clump}\lesssim5$, e.g.][]{ggf:11}. However, some of the simulated clumps would merge with each other in several orbital time-scales. In addition, it is argued that recent high-redshift observations tend to underestimate $N_{\rm clump}$ because of their insufficient spatial resolutions which erroneously combine a few clumps into a single one \citep[e.g.][]{bbs:16}. In low-redshift clumpy galaxies observed with higher resolutions, the numbers of clumps per galaxy are $4$--$18$ in the sample of \citet{f:17}. In the simulations shown above and in Appendix \ref{App}, we find that the turbulent velocity (velocity dispersion) of gas is approximately $\sim50~{\rm km~s^{-1}}$ in the rings, whereas the temperature $T=10^4~{\rm K}$ assumed in the simulations corresponds to the sound velocity of $c_{\rm snd}=15~{\rm km~s^{-1}}$. Hence, the gas on the rings is highly turbulent, and the thermal pressure hardly prevents the gas from fragmenting. This means that the assumed temperature does not artificially suppress the formation of small clumps on the rings.

It has been suggested that clumpy galaxies are formed through spontaneous instability of a galactic disc at high redshifts \citep[e.g.][]{n:98}. Fragmentation of CRGs can thus be an alternative path of clumpy galaxy formation although it does not account for all the clumpy galaxies.\footnote{Clumpy galaxies account for nearly half of the star-forming galaxies at $z\sim2$ \citep[e.g.][]{sok:16}.} \citet{eering:06} have discussed this scenario from their observations of ring galaxies with clumpy star formation at redshift $z\sim0.8$. \citet{mmr:08} argue that remnants of CRGs finally evolve into a population of giant low-surface brightness galaxies after their rings disappear. The clumps in our simulation also rotate in the diffuse disc. In the observations of \citet{eem:09}, a few galaxies hosting multiple clumps in underlying red discs at $z<1.5$ indicate nearly an order of magnitude lower surface densities in their inter-clump regions than other clumpy and spiral galaxies. Since such low-density discs are expected to be dynamically stable, the formation of their clumps cannot be explained by disc instabilities.

\section{Instability Analysis}
\label{ana}
\citet{iy:18} develop the linear perturbation analysis of \citet[][]{tti:16} to derive physical conditions for gravitational instability of a structure resembling a ring in a disc. The analysis considers azimuthal perturbations in an axisymmetric ring, in which the radial density profile is assumed to be a Gaussian function. \citet{iy:20} extend the analysis to include the stabilising effects by differential rotation and the vertical thickness of a ring. For a two-component ring consisting of gas and stars, the instability parameter for a perturbation with wavenumber $k$ is given as 
\begin{equation}
S_2 \equiv\frac{1}{\upi G k^2}\left[\frac{\Upsilon_{\rm g} f(kW_{\rm g})F(kh_{\rm g})}{\sigma_{\rm g}^2k^2 + \kappa_{\rm g}^2} + \frac{\Upsilon_{\rm s} f(kW_{\rm s})F(kh_{\rm s})}{\sigma_{\rm s}^2k^2 + \kappa_{\rm s}^2}\right]^{-1},
\label{Crit}
\end{equation}
where the suffixes `g' and `s' denote quantities of the gas and stellar components, $G$ is the gravitational constant, $W$ and $\Upsilon$ are the half-width and the mass of the ring per unit length (line-mass), $\sigma$ is the azimuthal velocity dispersion, and $f(kW)\equiv[K_0(kW)L_{-1}(kW) + K_1(kW)L_0(kW)]$ with $K_i$ and $L_i$ are the modified Bessel and Struve functions of order $i$. From \citet{t:64}, the thickness correction factor is given as $F(kh)=[1-\exp(-kh)]/(kh)$, where $h$ is the vertical thickness. If $S_2(k)<1$, the perturbation $k$ is expected to grow exponentially with time and therefore be unstable. The instability condition of the ring is given as $\min [S_2(k)] < 1$.

Local values of $\sigma$, $\Upsilon$, $W$, $\kappa$ and $h$ are computed for both gas and stars from the simulation outputs. We use vertically averaged values of these physical quantities by applying two-dimensional Gaussian smoothing of $0.2~{\rm kpc}$ to the gas and stellar  distributions. Then we make polar plots of the quantities as functions of $(R,\phi)$. For gas, $\sigma_{\rm g}^2 = c_{\rm snd}^2+\sigma_{\phi,{\rm g}}^2$, where $c_{\rm snd}$ and $\sigma_{\phi,{\rm g}}$ are the sound speed and dispersion of azimuthal turbulent velocities. For stars, $\sigma_{\rm s}$ is the dispersion of their azimuthal velocities. Thicknesses are $h_{\rm g}=(c_{\rm snd}^2+\sigma_{z,{\rm g}}^2)/(\upi G\Sigma_{\rm g})$ for gas and $h_{\rm s}=\sigma_{z,{\rm s}}^2/(\upi G\Sigma_{\rm s})$ for stars, where $\sigma_z$ and $\Sigma$ are vertical velocity dispersion and surface density. As in \citet{idm:16}, for each of gas and stars, we compute $\kappa$ from the mean rotation velocities $v_{\phi}$ as
\begin{equation}
  \kappa^2=2\frac{v_{\phi}}{R}\left(\frac{\mathrm{d}v_{\phi}}{\mathrm{d}R} + \frac{v_{\phi}}{R}\right),
  \label{kappa}
\end{equation} 
rather than from circular velocity estimated from a gravitational potential. This is because our linear perturbation analysis is based on the continuous and momentum equations described with the mean velocity $v_\phi$, rather than circular velocity. This treatment makes our analysis flexible to the possible change of the halo density distribution in the merging, and its accuracy is independent of the halo potential assumed in the simulations. In an expanding ring structure, $v_\phi$ would be able to characterise Coriolis forces more accurately than circular velocity. \citet{idm:16} have argued that using the mean velocity $v_\phi$ can make the linear analysis more accurate in a turbulent disc.

To measure the half-width $W$ of the ring, we use radial Gaussian fitting at a given $\phi$ on the polar plot of $\Sigma(R,\phi)$. The fitting function is defined as $\tilde{\Sigma}(R,\xi,\phi)=\Sigma(R,\phi)\exp[-\xi^2/2w^2]$, where $\xi$ represents the radial offset from $R$. While varying $w$, the fitting is iteratively performed in the range of $-1.55w<\xi<1.55w$. We search for a value of $w$ that minimises the goodness-of-fit $\chi^2$, and the half-width is defined as $W=1.55w$. With this definition, the estimated Gaussian surface density at the edge of the ring corresponds to 30 per cent of the peak value, i.e. $\tilde{\Sigma}(R,\pm W,\phi)=0.3\Sigma(R,\phi)$, and the line-mass is obtained as $\Upsilon(R,\phi)=1.44W\Sigma$. For computing $\kappa$, we calculate the local velocity gradient $\mathrm{d}v_{\phi}/\mathrm{d}R$ at $(R,\phi)$ by linear least-squares fitting for $v_{\phi}$ in the radial range of $-W<\xi<W$. 

\begin{figure}
  \includegraphics[bb=0 0 935 780, width=\hsize]{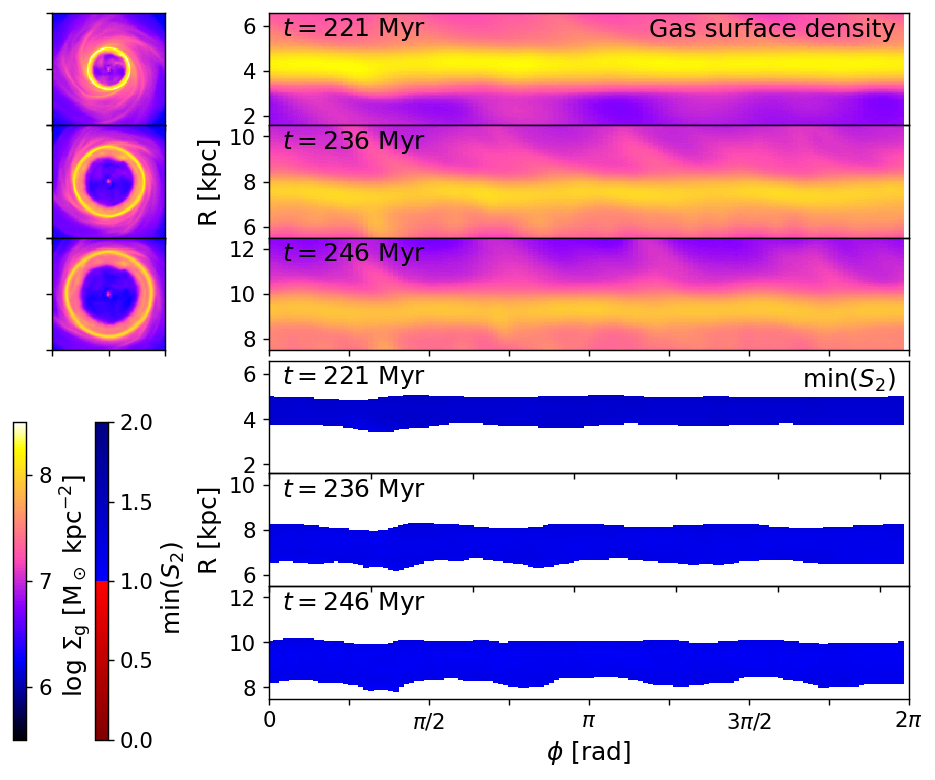}
\caption{Our instability analysis for the run with $f_{\rm g}=0.1$ at $t=221$, $236$ and $246~{\rm Myr}$. The top set of panels show surface densities of gas in Cartesian (left) and the polar (right) coordinates. The bottom set of panels show the instability parameters $\min(S_2)$ in the ring regions defined with $\log(\chi_{\rm g}^2+\chi_{\rm s}^2)<-0.3$.}
\label{Stable}
\end{figure}
\begin{figure}
  \includegraphics[bb=0 0 935 788, width=\hsize]{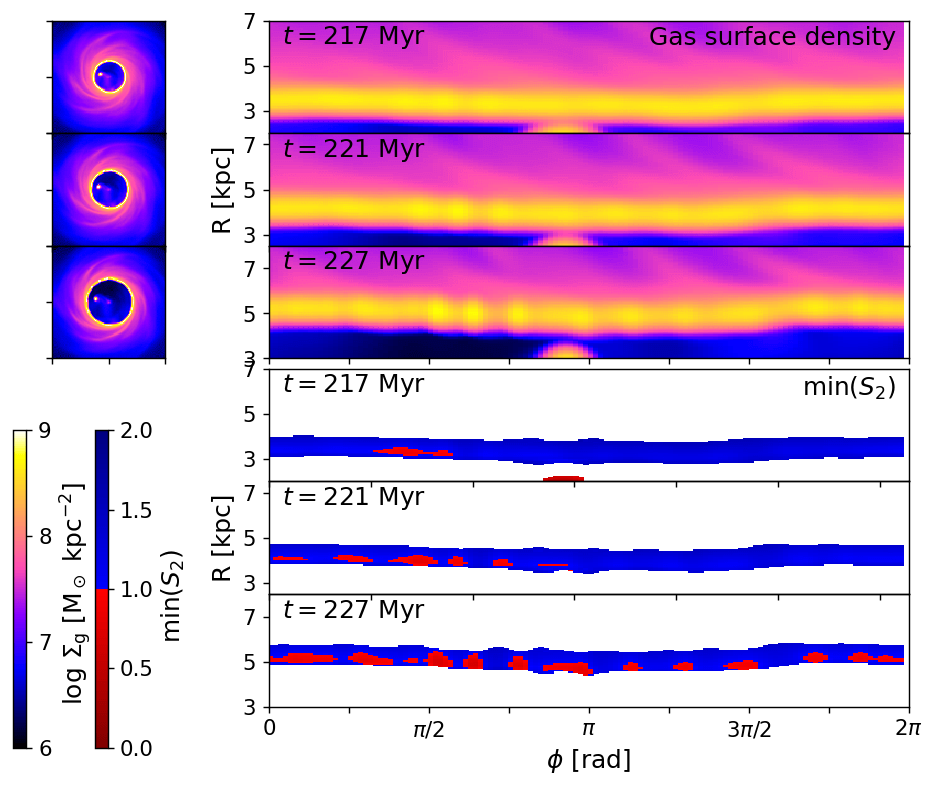}
\caption{Same as Fig. \ref{Stable} but for the run with $f_{\rm g}=0.15$ at $t=217$, $221$ and $227~{\rm Myr}$. At $t=217~{\rm Myr}$, the region indicating $S_2(k)<1$ at $(R,\phi)\simeq(2~{\rm kpc}, \upi~{\rm rad})$ is the off-centre nucleus formed by the impact of the collision.}
\label{Unstable}
\end{figure}
Fig. \ref{Stable} shows our instability analysis for the run with $f_{\rm g}=0.1$, where the CRG does not fragment. The bottom panels show that $\min(S_2)>1$, predicting stable states in all regions of the ring. Fig. \ref{Unstable} shows the results for the run with $f_{\rm g}=0.15$. Our analysis indicates $\min(S_2)<1$ in the regions at $\phi\simeq\upi/2$ at $t=217~{\rm Myr}$. In the following snapshots, other regions also indicate $\min(S_2)<1$. At $t=227~{\rm Myr}$, the ring fragments at $\phi\simeq\upi/2$ where $\min(S_2)$ is below unity. Our analysis can thus characterise the stability and the onset of ring fragmentation in our simulations. From the above results, we explain qualitatively that a ring fragments by its self-gravity when it is massive (large $\Upsilon$), thin (small $W$ and $h$) and/or cold (small $\sigma$) in a galaxy with small $\kappa$ (see equation \ref{Crit}). The same analysis applied to the additional set of simulations is presented in Appendix \ref{App} (see Figs. \ref{StableApp} and \ref{UnstableApp}).

\begin{figure}
  \includegraphics[bb=0 0 812 442, width=\hsize]{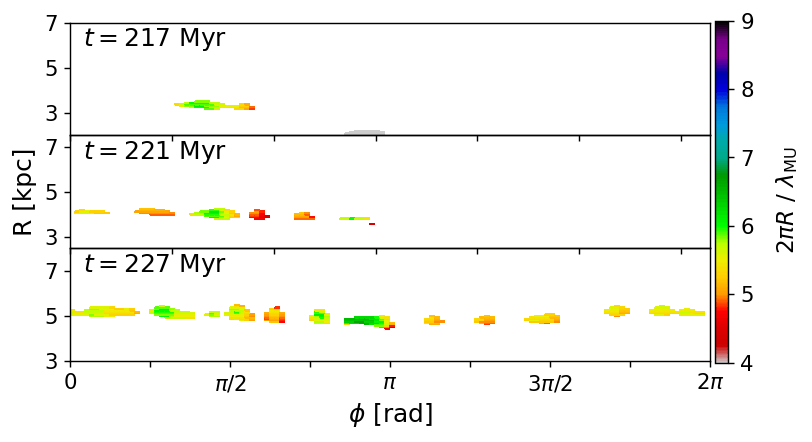}
\caption{The number of clumps predicted from the analysis as $N_{\rm clump}=2\upi R/\lambda_{\rm MU}$ in the fragmenting run with $f_{\rm g}=0.15$. Here, only the regions indicating the instability with $\min(S_2)<1$ are displayed according to the colour bar (see Fig. \ref{Unstable}).}
\label{Ncl_Fid}
\end{figure}
Our analysis can also be used to estimate the physical scales of the instability as $\lambda_{\rm MU}=2\pi/k_{\rm MU}$, where $k_{\rm MU}$ is the wavenumber that gives the minimum $S_2(k)$. Since the ring is almost circular before fragmenting, the number of clumps can be predicted as $N_{\rm clump}=2\upi R/\lambda_{\rm MU}$. Fig. \ref{Ncl_Fid} shows that $N_{\rm clump}=5$--$6$ is predicted in most of the unstable segments indicating $\min(S_2)<1$ in the ring. Although this is fewer than the actual clumps, $N_{\rm clump}=12$,\footnote{One of the 13 clumps is an off-centre nucleus formed by the impact of collision.} formed in the simulation, the prediction is consistent to within a factor of two. The under-prediction of $N_{\rm clump}$ implies that our analysis overestimates $\lambda_{\rm MU}$, and/or unstable perturbations may grow in a non-linear manner where an unstable perturbation shorter than $\lambda_{\rm MU}$ collapses first.\footnote{The linear perturbation theory expects that the most unstable perturbation with $\lambda_{\rm MU}$ collapses first \citep{bt:08}.} The under-prediction of $N_{\rm clump}$ is also seen in the analysis for our supplementary simulations (see Fig. \ref{Ncl_App} in Appendix \ref{App}). We further discuss the inconsistency of $N_{\rm clump}$ in Section \ref{discussion}.

\section{Swept-up approximation}
\label{Sweep}
Here we consider a simple model applicable to rings in the marginal states with $\min(S_2)\simeq1$, where their self-gravity is not negligible. If appropriate physical quantities are given, the physical model allows us to predict whether a CRG fragments before the merger. We assume that the impact of a merger forms a ring at $R$, which has swept up all the gas and stars inside $R$. The line-mass of the ring is then
\begin{equation}
\Upsilon=\frac{1}{2\upi R}\int^R_02\upi r\Sigma_{\rm IC}~{\mathrm{d}r},
\label{linemass}
\end{equation}
where $\Sigma_{\rm IC}$ is the surface density of the disc before the merger; i.e. the relaxed ICs. This model considers the ring to be a `material wave' in contrast with the density-wave theory of \citet{lt:76}. The ring expands while accumulating the gas and stars within the ring radius, obtaining the total mass of $M_{\rm ring}=2\pi R\Upsilon$. The accumulated material conserves its angular momentum, and thus the mean rotation velocity at $R$ is
\begin{equation}
v_\phi=\frac{1}{RM_{\rm ring}}\int^R_02\upi r^2v_{\phi,{\rm IC}}\Sigma_{\rm IC}~{\mathrm{d}r},
\label{velphi}
\end{equation}
where $v_{\phi,{\rm IC}}$ is the rotation velocity in the relaxed ICs. We approximate $\kappa=2v_\phi/R$ since self-gravitating structures such as rings have rigid rotation \citep{tti:16,iy:18}. Gas and stars migrate from $r$ to $R$ while their rotation velocities scale as $v_{\phi,{\rm IC}}(r/R)$. Since the ring sweeps up these with different velocities into the ring, it generates an azimuthal velocity dispersion, 
\begin{equation}
\sigma_{\rm A}^2=\frac{1}{M_{\rm ring}}\int^R_02\upi r\Sigma_{\rm IC}\left[v_\phi-v_{\phi,{\rm IC}}\frac{r}{R}\right]^2~{\mathrm{d}r}.
\label{sigmaA}
\end{equation}
We assume that the intrinsic dispersions of azimuthal and vertical velocities are approximately conserved while the ring expands,\footnote{Note that the accumulated materials can be in a new equilibrium with different velocity dispersions. We find, however, that values of $\sigma$ computed with this assumption are quite similar to those in the simulations.} and then the contribution is calculated as
\begin{equation}
\sigma_{\rm B}^2=\frac{1}{M_{\rm ring}}\int^R_02\upi r\Sigma_{\rm IC}\sigma_{\rm IC}^2~{\mathrm{d}r}.
\label{sigmaB}
\end{equation}
where $\sigma_{\rm IC}$ is the velocity dispersion in the relaxed ICs. We evaluate $\sigma_{\rm B}$ separately in the azimuthal and vertical directions. For the gas component, $\sigma_{\rm IC}$ includes the contributions from the sound velocity. Then, the azimuthal and vertical velocity dispersions in the ring are computed as $\sigma_\phi^2=\sigma_{\rm A}^2+\sigma_{{\rm B},\phi}^2$ and $\sigma_z^2=\sigma_{{\rm B},\phi}^2$, respectively. The above equations (\ref{linemass}--\ref{sigmaB}) set the lower limit of the integration to be $R=0$ where there can be a bulge, and a companion would not penetrate the exact centre. However, the contribution from the central regions is expected to be small in the integration.

Finally, we need to evaluate $W$ for the analysis to be completed. In our simulations, we find that the ring widths $W$ are nearly constant with time until they fragment or reach $R\simeq7~{\rm kpc}$: $W_{\rm s}\simeq2~{\rm kpc}$ and $W_{\rm g}\simeq0.5~{\rm kpc}$ in both runs. Hence, we approximate $W$ as constant in the radial range from $R\sim2$ to $7~{\rm kpc}$. By assuming a Gaussian density distribution, the surface density at the crest of the ring is given as $\Sigma=\Upsilon/(1.44W)$.

\begin{figure}
  \includegraphics[bb=0 0 1121 698, width=\hsize]{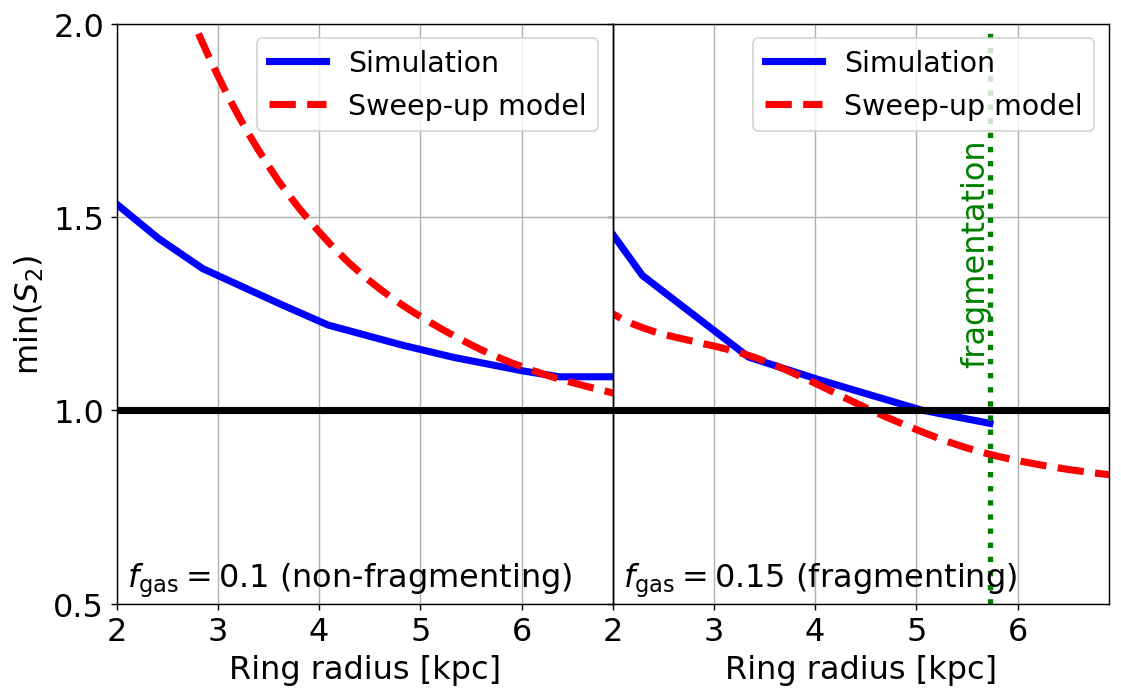}
\caption{The instability parameters $\min(S_2)$ as functions of ring radius $R$. The left and right panels show the results for the cases of $f_{\rm g}=0.1$ and $0.15$. For the simulations, we compute the median values of $S_2$ measured along the crests of the simulated rings. In the right panel, the simulated CRG completely fragments at $R\simeq5.7~{\rm kpc}$ (the vertical green dotted line), and therefore $S_2$ is not measured at the larger radius.}
\label{sweepupmodel}
\end{figure}
Using the above model and the relaxed ICs, we compute $\Upsilon$, $\sigma$, $\kappa$ and $h$ as functions of ring radius $R$ for each of the gas and stellar components. We substitute the quantities obtained from the model into equation (\ref{Crit}) and compute $\min(S_2)$. Fig. \ref{sweepupmodel} compares the results with the simulations. For the run with $f_{\rm g}=0.1$ (left), the swept-up model and the simulation are consistent in terms of the stability indicated by $\min(S_2)>1$ in $R\lesssim7~{\rm kpc}$ where the assumption of the constant $W$ is valid. Although the model indicates larger $\min(S_2)$ than the simulation at small $R$, the values are close to the simulation result at large $R$ and do not fall below unity. In the run with $f_{\rm g}=0.15$ (right), the swept-up model indicates $\min(S_2)$ consistent with or slightly below the simulation result. It is noteworthy that the swept-up model predicts the unstable states for the ring with $\min(S_2)<1$ at $R=4.5~{\rm kpc}$, and this is in agreement with the simulation. These results of our toy model support the ability of our linear perturbation analysis to characterise the dynamical instability of CRGs, and the self-gravitating rings in our simulations can be approximated as material waves accumulating gas and stars in the discs.  

As we noted above, the swept-up model indicates the deviation of $\min(S_2)$ from those in the simulations when ring radii are small. It is possibly because of contraction of the discs which is caused by the companions in the early phases of the mergers. A companion passing through a disc centre attracts stars and gas inwards in the disc, and therefore the disc becomes temporarily dense and compact during the merger \citep{lt:76,bt:08}. Although the disc expands as the ring propagates outwards, the disc still has a higher surface density than its isolated state when the ring radius is small. Hence, in the early phases of the mergers, line-masses $\Upsilon$ in the simulations are larger than those predicted from the swept-up model; this effect lowers $S_2$. In addition, the disc contracts while conserving angular momentum, and therefore its rotation velocity increases in the contraction. Hence, the simulated rings have larger $\kappa$ than the model prediction at small radii; this effect raises $S_2$. Although these effects counteract each other in computing $S_2$, they do not cancel completely. These are not taken into account in our swept-up model. After the rings expand to large radii, we find that systematic differences of $\Upsilon$ and $\kappa$ are not significantly large between the swept-up model and the actual rings in the simulations, and therefore $\min(S_2)$ becomes consistent. We also show the same results but for the supplementary runs in Appendix \ref{App} (see Figs. \ref{sweepupmodelApp}).

\section{A high-redshift CRG}
\label{highz}
R5519 has been discovered at redshift $z=2.19$ and is the most distant CRG ever observed \citep{yel:20}. Here we adopt our instability analysis to this high-redshift CRG. The observations measure various physical properties of R5519 as follows. The CRG has a total star formation rate (SFR) of $38~{\rm M_\odot~yr^{-1}}$ and stellar mass of $\log(M_{\rm s}/{\rm M_\odot})=10.78$. It is estimated that $45$--$70$ per cent of the total SFR resides in the ring. The ring is nearly circular and has a radius of $R=5.1~{\rm kpc}$, a width of $3.7~{\rm kpc}$, and an estimated rotation velocity of $v_\phi=90~{\rm km~s^{-1}}$.
 
Using these observed quantities, we consider two extreme cases for R5519: the minimum and maximum ring models. The former (latter) model assumes the least (most) massive ring and provides the upper (lower) limit of $S_2$. In the minimum ring model, we ignore the stellar component and assume that the ring has $45$ per cent of the total SFR of the galaxy. The averaged SFR surface density in the ring is estimated to be $\overline{\Sigma}_{\rm SFR}=0.14~{\rm M_\odot~yr^{-1}~kpc^{-2}}$, and this is converted to the averaged gas surface density of $\overline{\Sigma}_{\rm g}=9.3\times10^7~{\rm M_\odot~kpc^{-2}}$ through the star formation law of \citet{k:98}. Although the measurements of the ring width by \citet{yel:20} are not exactly compatible with our definition of $W$, we take the width to be $2W=3.7~{\rm kpc}$ in our analysis. Adopting the Gaussian distribution for the ring, the surface density at the crest is calculated to be $\Sigma_{\rm g}=1.41\overline{\Sigma}_{\rm g}=1.3\times10^8~{\rm M_\odot~kpc^{-2}}$. We assume rigid rotation for the ring: $\kappa=2v_{\phi}/R$. Because the velocity dispersion $\sigma$ of the ring is not measured, we compute $\min(S_2)$ as a function of $\sigma$. We assume the velocity dispersion is isotropic: $\sigma_\phi=\sigma_z$. From equation (\ref{Crit}), $\min(S_2)$ monotonically increases with $\sigma$. We find that $\min(S_2)<1$ when $\sigma<27~{\rm km~s^{-1}}$. This model ignores the stellar component and uses the lower limit of $\overline{\Sigma}_{\rm SFR}$ to provide the upper limit, and R5199 is therefore expected to be unstable if $\sigma<27~{\rm km~s^{-1}}$.

The maximum ring model takes into account the stellar component and assumes that the ring has $70$ per cent of the total SFR. Since the width and mass of the stellar ring are not measured, we assume $W_{\rm s}=W_{\rm g}$ and that $70$ per cent of the total stellar mass of the galaxy resides in the ring. We consider that $v_\phi$ is the same between the gas and stellar components. The free parameter $\sigma$ is again assumed to be isotropic, and the stellar component shares the same $\sigma$ with the gas. Assigning $W$ and $\sigma$ of the gas to the stars can significantly underestimate $S_2$ since the stars are thought to have larger $W$ and $\sigma$ than the gas. Moreover, assuming the maximum $\overline{\Sigma}_{\rm SFR}$ for the ring also results in relatively low $S_2$. Thus, $\min(S_2)$ in this model gives a lower limit. We find that $\min(S_2)<1$ when $\sigma<105~{\rm km~s^{-1}}$ in this model. This means that R5199 is expected to be stable if $\sigma>105~{\rm km~s^{-1}}$.

In the range between $\sigma=27$ and $105~{\rm km~s^{-1}}$, there remains the possibility of dynamical instability. Turbulent velocities of star-forming discs at $z\sim2$ are observed to range from $\sigma\sim20$--$130~{\rm km~s^{-1}}$ with a median value of $\sim50~{\rm km~s^{-1}}$ \citep{yrg:17}. If the progenitor of R5199 is such a typical galaxy with $\sigma\sim50~{\rm km~s^{-1}}$, it is possible that the ring instability operates, and the high-redshift CRG would evolve into a clumpy galaxy.

Although CRGs are rare and account for only $\sim0.01$ per cent of galaxies in the local Universe \citep{mnp:09}, they are expected to be more abundant at higher redshifts since mergers are more frequent then. \citet{yel:20} argue, however, that CRGs at $z\sim2$ are as rare as in the local Universe, and this is an open question of high-redshift galaxies. They show that R5199 has a significantly higher surface brightness than the local CRGs, and the Cartwheel Galaxy would be too faint to observe if it were at the same redshift as R5199. If a high-redshift CRG is unstable, its fragmentation can drive active star formation and may elevate the brightness of the ring to an observable level. Not all CRGs would be unstable at high redshifts. Therefore, if the current high-redshift observations are limited to such unstable CRGs that can be bright enough, the instability of rings may explain the rarity of high-redshift CRGs.

\section{Discussion and conclusions}
\label{discussion}
By computing the instability parameter $\min(S_2)$ given in equation (\ref{Crit}), our linear perturbation analysis can describe the (un-)occurrence of fragmentation of the CRGs in our simulations. Although the density-wave theory of \citet{lt:76} ignores self-gravity of rings, our analysis including self-gravity can describe the dynamical states of massive CRGs and characterise their fragmenting instability. The analysis using equation (\ref{Crit}) can be adopted not only to simulation snapshots but also to observational data if the relevant physical quantities are available, as demonstrated in Section \ref{highz} and \citet[][for spiral arms]{itm:21}. In Fig. \ref{sweepupmodel}, our swept-up model computes the evolution of $\min(S_2)$ from the initial conditions of the simulations. Despite its simplicity, the model can capture whether the simulated CRGs fragment or not although contraction of the discs affects the results in the early phases of ring fragmentation. However, the toy model is not completely stand-alone but uses $W$ measured in the actual simulations. Our simulations seem to tentatively show that the ring widths $W_{\rm g}$ and $W_{\rm s}$ do not vary largely during the early phases of ring expansion although it is not justified. If we can approximate $W$ from a progenitor disc without running a simulation, our swept-up model can predict the evolution of a CRG before a head-on merger takes place. It may enable us to estimate, from observations, the fraction of disc galaxies that can fragment by strong perturbations.

Figs. \ref{Unstable} and \ref{sweepupmodel} demonstrate that the ring fragments when it reaches $\min(S_2)<1$ in our simulation and the toy model. These figures show, however, that $\min(S_2)$ is not necessarily constant with time. This is mainly because $\Upsilon$ and $\kappa$ vary as the ring expands. Therefore, even if a ring indicates $\min(S_2)>1$ at a certain time, the instability may operate later at an outer radius. However, if the relevant physical quantities can be obtained from observations, the subsequent evolution of $\min(S_2)$ can be predicted using the swept-up model. The toy model assumes the material in a galaxy is accumulated into a ring, and thus is able to estimate $\min(S_2)$ given by equation (\ref{Crit}) at an arbitrary ring radius. Hence our analysis can predict the fate of self-gravitating CRGs.

For R5199, the available information is still limited. Especially, determining turbulent velocity $\sigma_{\rm g}$ with spectroscopic observations would be a key to improve our estimation of $\min(S_2)$; the properties of the stellar component are also important. As high-redshift disc galaxies tend to form giant clumps due to their highly gas-rich nature, CRGs may be prone to fragmentation at high redshifts. We expect that our instability analysis is essential for understanding high-redshift CRGs such as R5519.

In Fig. \ref{Ncl_Fid}, however, it appears that our analysis underpredicts the number of fragments of the unstable CRG by a factor of $\sim2$. Here, we mention a few possible reasons for the underestimate for $N_{\rm clump}$. First, our analysis assumes the Gaussian density distribution for a ring, and it is reflected in equation (\ref{Crit}) via the function $f(kW)$. This assumption has not been justified and may not be accurate in CRGs. A possible improvement may be to relax the Gaussian assumption by describing $f(kW)$ in a more flexible form with additional parameters for the density distribution. A second possibility may be the limited accuracy of the linear perturbation theory, which is derived from the linearised equations of continuity and momenta \citep[see][]{iy:18,tti:16}. Although the linear analysis can characterise the (un-)occurrence of fragmentation in the simulated CRGs, it may not be accurate enough to describe the detailed behaviour of unstable perturbations, such as the physical scale of fragmentation $\lambda_{\rm MU}$. Similar inaccuracy has also been reported in \citet{iy:18} for the instability of spiral arms, where the analysis appears to overestimate the growth time-scales of unstable perturbations by a factor of a few, although the criterion for $\min(S_2)$ describe well fragmentation of spiral arms \citep[see also][]{itm:21}.

\section*{Acknowledgements}
We are grateful to an anonymous reviewer for his/her useful comments. We thank Volker Springel for kindly providing the simulation code {\sc Arepo}. This study was supported by World Premier International Research Center Initiative (WPI), MEXT, Japan. The simulations and numerical analysis presented in this paper were carried out on Cray XC50 and the analysis servers at Center for Computational Astrophysics, National Astronomical Observatory of Japan.

\section*{Data Availability}
The data underlying this article will be shared on reasonable request to the corresponding author.



\bibliographystyle{mnras}





\appendix
\section{Results for simulations with other initial conditions}
\label{App}
To test our analysis further, we run similar simulations using different initial conditions with a smaller disc mass and a higher $f_{\rm g}$ than in the fiducial cases. The total disc mass of the main galaxy is set to $2.8\times10^{10}~{\rm M_\odot}$, and the disc scale radius and height are the same as the fiducial ones. In the ICs, the original gas fraction is $f_{\rm g}=0.2$, and the temperature is the same: $10^4~{\rm K}$. The disc is modelled with $10^6$ gas cells and $5\times10^6$ stellar particles. The halo potential is unchanged, and the initial velocity distribution of the gas and stars is determined by the minimum $Q$ value of $1.3$. This IC is evolved in isolation for $1~{\rm Gyr}$ in the same relaxation process (see Section \ref{sims}), and we create two relaxed ICs with $f_{\rm g}=0.2$ and $0.3$. The merging companion is represented with the same potential and mass, which is placed initially at $52.5~{\rm kpc}$ above the disc centre with the escape velocity of $174~{\rm km~s^{-1}}$.

\begin{figure*}
  \includegraphics[bb=0 0 1542 384, width=\hsize]{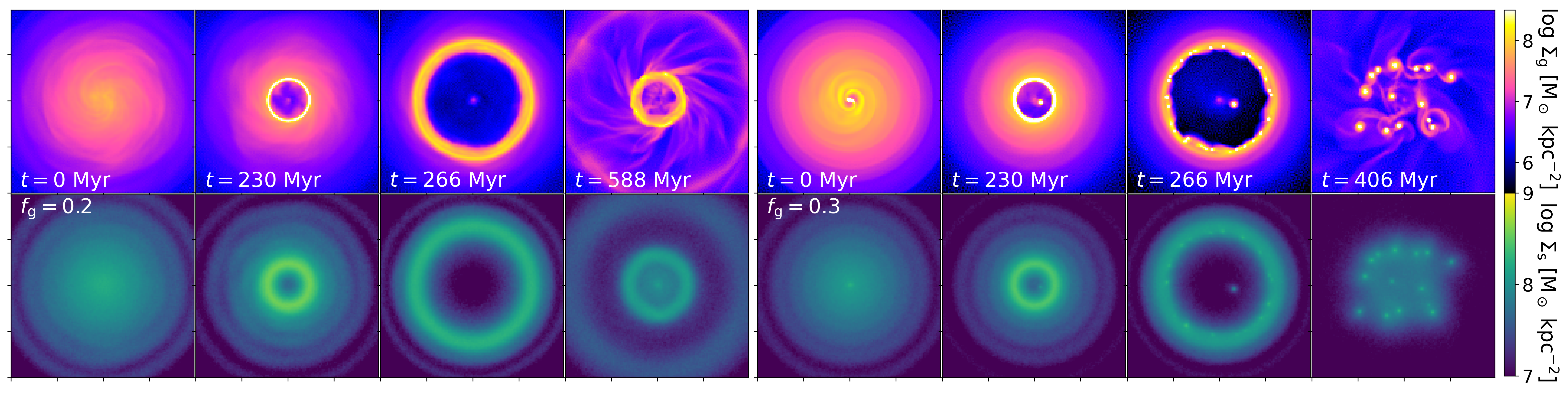}
\caption{Same as Fig. \ref{snaps} but for the supplementary simulations with the smaller total disc mass of $2.8\times10^{10}~{\rm M_\odot}$. The gas fractions are $f_{\rm g}=0.2$ and $0.3$ in the left and right sets of panels. The companions pass through the disc centres at $t=208~{\rm Myr}$.}
\label{snapsApp}
\end{figure*}
\begin{figure}
  \includegraphics[bb=0 0 935 780, width=\hsize]{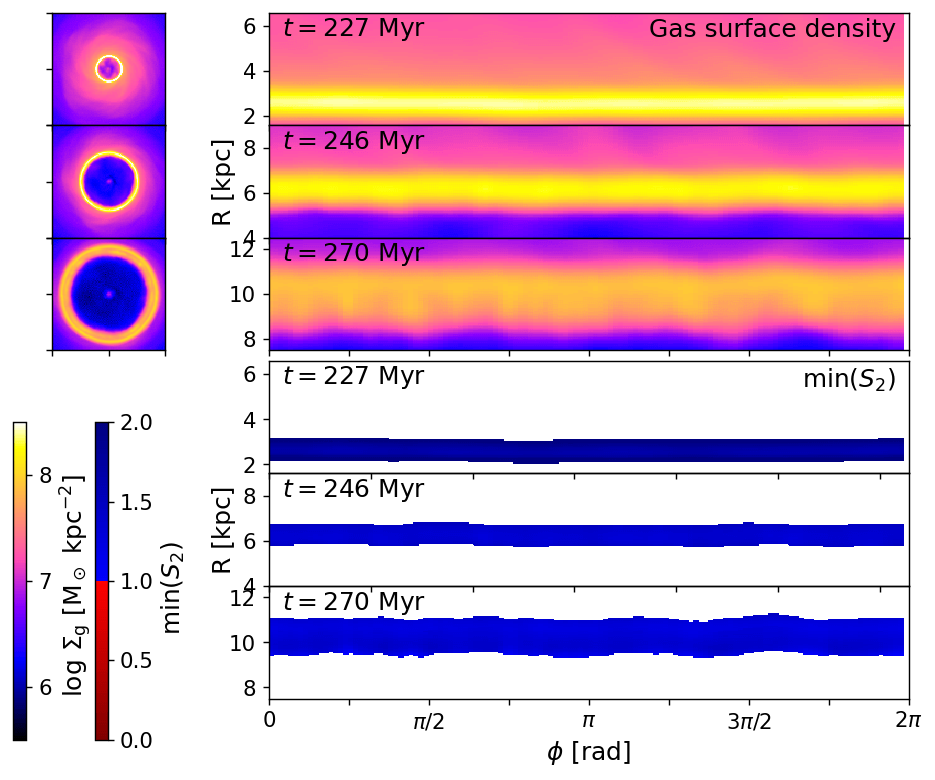}
\caption{Same as Fig. \ref{Stable} but for the non-fragmenting run with the smaller disc mass and $f_{\rm g}=0.2$ at $t=227$, $246$ and $270~{\rm Myr}$.}
\label{StableApp}
\end{figure}
\begin{figure}
  \includegraphics[bb=0 0 935 780, width=\hsize]{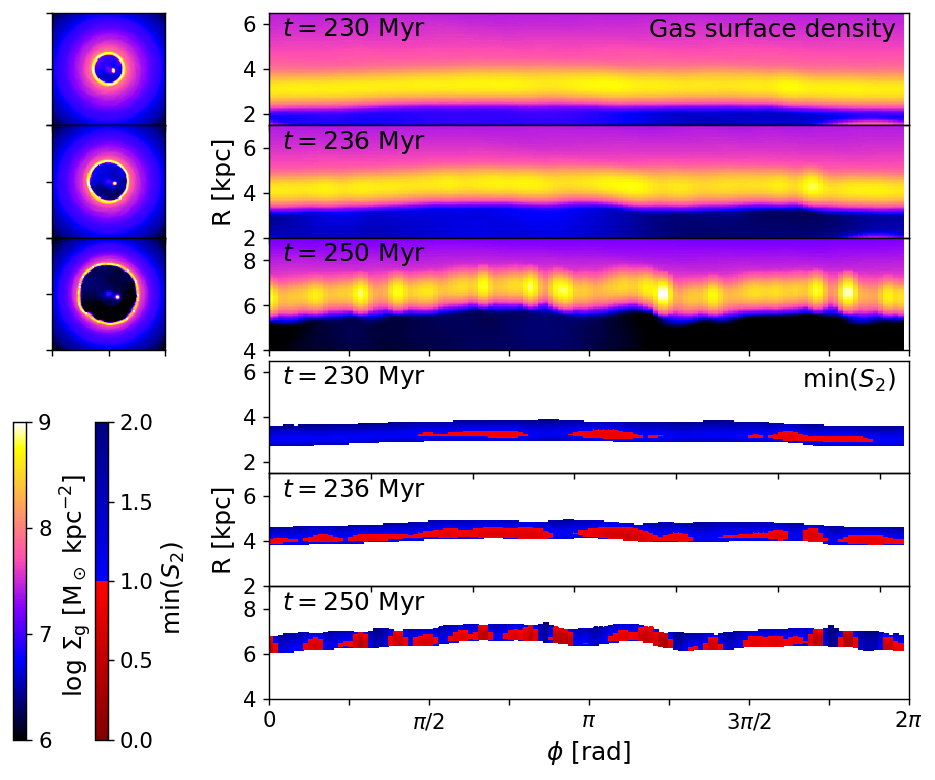}
\caption{Same as Fig. \ref{StableApp} but for the supplementary run with $f_{\rm g}=0.3$ at $t=230$, $236$ and $250~{\rm Myr}$.}
\label{UnstableApp}
\end{figure}
Fig. \ref{snapsApp} shows the face-on density distributions in the simulations. The CRG does not fragment in the supplementary run with $f_{\rm g}=0.2$, whereas the ring fragments into massive clumps at $R\sim6~{\rm kpc}$ when $f_{\rm g}=0.3$. Figs. \ref{StableApp} and \ref{UnstableApp} illustrate polar maps of gas surface densities and the instability parameters $\min(S_2)$ for the runs with $f_{\rm g}=0.2$ and $0.3$, respectively. The instability parameters indicate high values of $\min(S_2)>1$ in the non-fragmenting run with $f_{\rm g}=0.2$, whereas those indicate low values of $\min(S_2)<1$ in advance of the fragmentation in the run with $f_{\rm g}=0.3$. These results are consistent with those shown in Figs \ref{Stable} and \ref{Unstable}, confirming the ability of our analysis to characterise the fragmentation.

\begin{figure}
  \includegraphics[bb=0 0 812 442, width=\hsize]{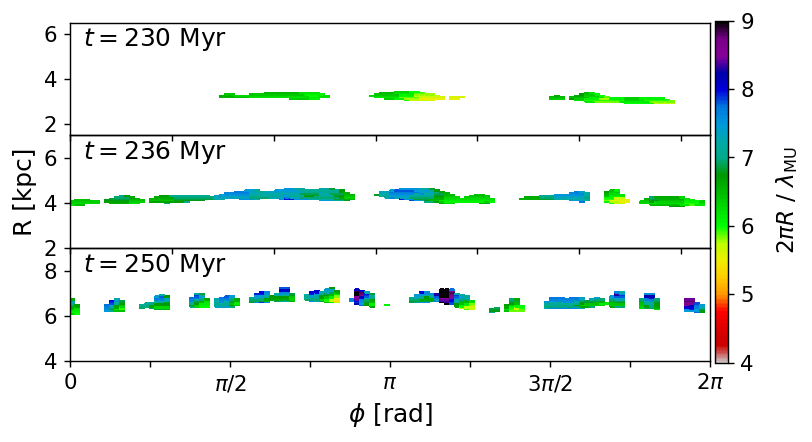}
\caption{Same as Fig. \ref{Ncl_Fid} but for the supplementary run with $f_{\rm g}=0.3$ at $t=230$, $236$ and $250~{\rm Myr}$.}
\label{Ncl_App}
\end{figure}
In the supplementary run with $f_{\rm g}=0.3$, the unstable CRG fragments into 13 clumps, and we find that their averaged gas mass is $3.6\times10^8~{\rm M_\odot}$ at $t=406~{\rm Myr}$ (the rightmost panel of Fig. \ref{snapsApp}). Fig. \ref{Ncl_App} shows the number of clumps predicted from the analysis as $N_{\rm clump}=2\upi R/\lambda_{\rm MU}$, and we obtain $N_{\rm clump}=6$--$7$ before the fragmentation. The predicted $N_{\rm clump}$ is thus nearly half of the actual number of the clumps formed in the run with $f_{\rm g}=0.3$. This under-prediction is also consistent with the result in the fiducial run with $f_{\rm g}=0.15$ in Section \ref{ana}.

\begin{figure}
  \includegraphics[bb=0 0 1121 698, width=\hsize]{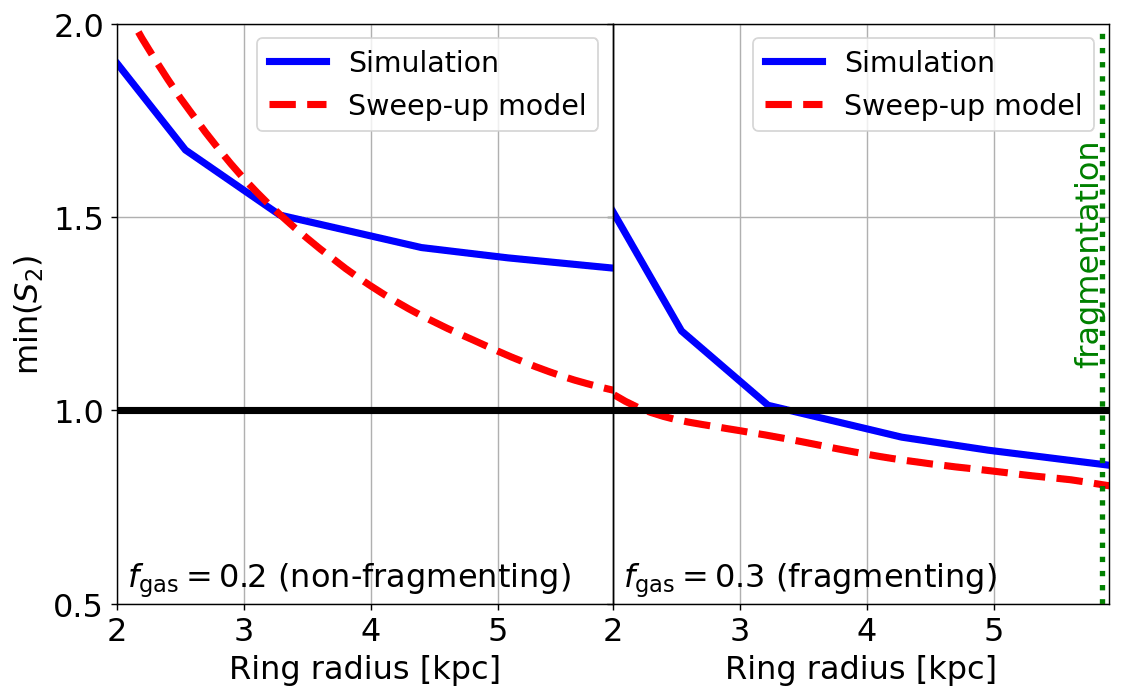}
\caption{Same as Fig. \ref{sweepupmodel} but for the supplementary runs with $f_{\rm g}=0.2$ (left) and $0.3$ (right). In the fragmenting run with $f_{\rm g}=0.3$, the ring fragments at $R\simeq6~{\rm kpc}$.}
\label{sweepupmodelApp}
\end{figure}
Fig. \ref{sweepupmodelApp} shows our results of the swept-up approximation described in Section \ref{Sweep} for the supplementary runs with $f_{\rm g}=0.2$ (left) and $0.3$ (right). As was done in Section \ref{Sweep}, we need to take the half widths of the rings from the simulation results. We find that the widths are nearly constant at $W_{\rm g}=0.5$ and $W_{\rm s}=1.5~{\rm kpc}$ until the rings expand to $R\sim6~{\rm kpc}$ in both runs.\footnote{Outside this radius, the half widths increase as the rings expand. This radius $R\sim6~{\rm kpc}$ sets the outer limit of the radial range where our swept-up model is applicable.} In the non-fragmenting run with $f_{\rm g}=0.2$, although the swept-up model appears to underestimate $\min(S_2)$ in $R\gtrsim4~{\rm kpc}$, the model predicts stable states with $\min(S_2)>1$ and is consistent with the absence of fragmentation in the simulation. In the fragmenting run with $f_{\rm g}=0.3$, the swept-up model underestimates $\min(S_2)$ in all radii until the ring fragments. The predicted values are, however, close to those obtained in the simulation in our radii, and the low values of $\min(S_2)<1$ are consistent with the actual fragmentation in the simulation.

The results shown here demonstrate the robustness of our analysis adopted to the CRG simulations. Our analysis can describe the (in)stability of the simulated CRGs despite the differences of the simulations in the total disc mass and $f_{\rm g}$. The accuracy of our model appears to be similar to that in Sections \ref{ana} and \ref{Sweep} although the prediction underestimates $N_{\rm clump}$. In the fiducial runs described in Section \ref{sims}, we also confirm that our results do not change with the impact velocity of the companion in the range between $1.5$ times lower and higher values than the escape velocity.


\bsp	
\label{lastpage}
\end{document}